\documentclass[aps, a4paper,superscriptaddress, nofootinbib, showpacs, twocolumn, 10pt]{revtex4}

\usepackage{amsmath,amssymb,bm,natbib}
\usepackage{graphicx}
\usepackage{slashed}

\def\be{\begin{equation}}
\def\ee{\end{equation}}
\def\bea{\begin{eqnarray}}
\def\eea{\end{eqnarray}}

\def\reff#1{\ref{#1}}

\def\eq#1{Eq.~(\reff{#1})}
\def\eqs#1#2{Eqs.~(\reff{#1}) and (\reff{#2})}
\def\fig#1{Fig.~\reff{#1}}

\def\tab#1{Table~\reff{#1}}

\begin{document}

\title{Model-independent description of  $B\to \pi l\nu$
decays  and a determination of $|V_{ub}|$}
\author{Claude Bourrely}
\affiliation{Centre de Physique Th\'eorique, UMR 6207, CNRS-Luminy, Case 907 
F-13288 Marseille Cedex 9, France}
\author{Irinel Caprini}
\affiliation{National Institute of Physics and Nuclear Engineering, Bucharest POB MG-6, R-077125 Romania} 
\author{ Laurent Lellouch}
\affiliation{Centre de Physique Th\'eorique, UMR 6207, CNRS-Luminy, Case 907 
F-13288 Marseille Cedex 9, France} 

\begin{abstract} 
  We propose a new parametrization of the $B\to\pi$ vector form
  factor, $f_+(q^2)$, as an expansion in powers of a conformal mapping
  variable, which satisfies unitarity, analyticity and perturbative
  QCD scaling. The unitarity constraint is used also for defining the
  systematic error of the expansion. We fit with the new
  parametrization the available experimental and theoretical
  information on exclusive $B\to\pi l\nu$ decays, making a
  conservative estimate of the effects of correlations in the
  systematic and statistical errors of the lattice results. With four
  parameters to describe $f_+(q^2)$, the systematic error is
  negligible in the whole semileptonic region. We also obtain
  $|V_{ub}|=(3.54 \pm 0.30) \times 10^{-3}$ where, in our approach,
  the uncertainty is predominantly statistical.
\end{abstract}

\maketitle
\section{Introduction}\label{sec:introduction}

As shown recently \cite{Arnesen:2005ez,Flynn:2006vr,Flynn:2007qd,Flynn:2007ii}, the extraction of
$|V_{ub}|$ from the exclusive semileptonic $B\to \pi$ decays has become
competitive with determinations from inclusive decays. The exclusive decay
approach requires the theoretical description of the matrix element
\bea\label{def} \langle\pi(p_\pi)|\bar{u}\gamma_\mu b |
B(p_\pi+q)\rangle && \!\!\!\!=  \nonumber \\ && \hspace{-2cm}\left(2 p_{\pi\mu} + q_\mu  
  -q_\mu\frac{m_B^2-m_\pi^2}{q^2}\right)  f_+(q^2)  \nonumber \\&& \hspace{-2cm} +\,
q_\mu\frac{m_B^2-m_\pi^2}{q^2} f_0(q^2), \eea
where $q$ is the momentum of the lepton pair and $f_+(q^2)$ and
$f_0(q^2)$ denote the vector and scalar form factors, respectively.
For light leptons, $l=e,\mu$, only the vector form factor
contributes to the spectrum 
\be\label{spectrum} \frac{d\Gamma}{dq^2} (\bar B^0\to\pi^+
l^-\bar\nu_\ell)=\frac{G_F^2|V_{ub}|^2}{192 \pi^3 m_B^3}
\,\lambda^{3/2}(q^2)|f_+(q^2)|^2, \ee where
$\lambda(q^2)=(m_B^2+m_\pi^2-q^2)^2 -4m_B^2m_\pi^2$ is $4m_B^2$ times
the pion three-momentum squared in the $B$-meson rest frame.  The
physical range of semileptonic decays is $0\le q^2 \le t_-$, with
$t_-=(m_{B^0}-m_{\pi^+})^2=26.42\,\mbox{GeV}^2$.

The $q^2$-spectrum of $B\to \pi \ell\nu$ decays has been measured with
increasing precision by the CLEO \cite{Athar:2003yg, Adam:2007},  Belle 
\cite{Hokuue:2006nr}, and  BaBar
collaborations \cite{Aubert:2006, Aubert:2007px}.  The accuracy of theoretical
calculations of the form factors is also continuously improving.  The
calculations are based either on QCD light-cone sum-rules (LCSR),
which provide reliable determinations at small $q^2$
\cite{Khodjamirian:1997ub,Ball:2004rg,Duplancic:2008ix}, or on lattice
simulations, which give accurate results at larger values of $q^2$
\cite{Okamoto:2004xg,Okamoto:2005zg,VandeWater:2006zz,Mackenzie:2005wu,Dalgic:2006dt, Bailey:2008wp}.

From general principles of quantum field theory it is known that the
vector form factor $f_+(q^2)$ is a real analytic function in the
complex $q^2$-plane cut for $q^2\ge t_+$, where
$\sqrt{t_+}=(m_{B^0}+m_{\pi^+})=5.419\, \mbox{GeV}$ is the $B\pi$
threshold.  Angular momentum conservation imposes the behavior
$\mbox{Im}\, f_+(q^2)\sim (q^2-t_+)^{3/2}$ near the threshold. In
addition, $f_+(q^2)$ has a pole below the branch point, at $q^2
=m_{B^*}^2$ ($m_{B^*}=5.325 \, \mbox{GeV}$). As shown in
\cite{Lellouch:1995yv}, unitarity applied to a certain QCD correlator
provides a constraint on the magnitude of $f_+(q^2)$ along the cut,
while scaling in perturbative QCD requires a power-law fall-off, like
$1/q^2$, up to logarithmic corrections \cite{Lepage:1980fj,
  Akhoury:1993uw}.

To maximize the usefulness of this information about $B\to \pi
\ell\nu$ decays and, for instance, determine $|V_{ub}|$, it is crucial
to have a parametrization of
$f_+(q^2)$,  which satisfies the above theoretical requirements,  and whose
associated systematic error is quantifiable and small compared to
experimental and other theoretical errors. The purpose of the present work is to provide such a parametrization.
In  Sec. \ref{sec:other}, we  discuss the recent models
proposed in the literature,  and show that they do not satisfy all of
the constraints mentioned above. In Sec. \ref{sec:new}, we propose a simple analytic
parametrization of $f_+(q^2)$, which combines the pole factorization
with an expansion in powers of a conformal mapping variable, and in Sec. \ref{sec:unitarity}, we express the unitarity constraints in terms of the coefficients of this expansion.  By
fitting the experimental differential decay rate $B\to\pi l\nu$ and
the values of $f_+(q^2)$ calculated from LCSR and the lattice, we
obtain in Sec. \ref{sec:best} a model-independent representation of the form factor
$f_+(q^2)$ in the physical region. The procedure also yields a precise value of $|V_{ub}|$, given in Sec. \ref{sec:Vub}. 
\section{Recent parametrizations of the form factor}\label{sec:other}

A comparison of the various parametrizations proposed in the
literature was done recently in \cite{Ball:2006jz}.  As shown there,
the simple expressions with a pole at $q^2=m_{B^*}^2$, proposed in
\cite{Ball:2004rg,Becirevic:1999kt}, cannot provide an accurate
description of the form factor in the whole physical domain. Thus,
more systematic expansions, which incorporate the constraints of
analyticity and unitarity, have been considered.

A first type of parametrization is inspired by the technique of
unitarity bounds proposed by Okubo \cite{Okubo:1971my,Okubo:1971jf}
and applied subsequently to various semileptonic form factors
\cite{Bourrely:1980gp,Caprini:1997mu,Bourrely:2005hp}. They are
obtained by exploiting the analyticity and positivity properties of
vacuum polarization functions of the type  $\Pi_{\mu\nu}(q){=}i\int
d^4x e^{iq\cdot x}$ $\langle 0\vert T\{ J_\mu(x)$
$J_\nu^\dagger(0)\}\vert 0\rangle$, where $J_\mu=\bar u\gamma^\mu b$
here. In this approach, the form factor can be expanded as
\cite{Boyd:1995cf,Arnesen:2005ez}
\be\label{AGRS} f_+(q^2)= \frac{1}{B(q^2) \phi(q^2,
  t_0)}\,\sum\limits_{n\ge 0} a_n(t_0) z^n \,, \ee
where $a_n(t_0)$
are real coefficients and $z\equiv z(q^2, t_0)$ is the function
\begin{equation}\label{z}
z(q^2, t_0)=\frac{\sqrt{t_+-q^2}-\sqrt{t_+-t_0}}{\sqrt{t_+-q^2}+ \sqrt{t_+-t_0}},
\end{equation}
which maps the $q^2$-plane cut for $q^2\ge t_+$ onto the disk $|z(q^2,
t_0) |<1$ in the $z$ complex plane, such that $z(t_+, t_0)=-1$ and
$z(\infty, t_0)=1$. The parameter $t_0<t_+$, which is arbitrary,
determines the point $q^2$ mapped onto the origin in the $z$ plane,
i.e. $z(t_0, t_0)=0$.

The function  $B(q^2)$ is the Blaschke factor
\be\label{Blas}
B(q^2)=\frac{z(q^2, t_0)-z(m_{B^*}^2, t_0)}{1- z(q^2, t_0) z(m_{B^*}^2, t_0)}= z(q^2, m_{B^*}^2 )\,,
\ee
which accounts for the pole at $q^2=m_{B^*}^2$.  By construction
$|B(q^2)|=1$ for  $q^2\ge t_+$. 

The last factor in (\ref{AGRS}), the outer function  $ \phi(q^2, t_0)$,
has the expression
\bea\label{phi}
 \phi(q^2, t_0)= &&\!\!\! \!\sqrt{\frac{1}{32 \pi \chi_{1^-}(0)}  } (\sqrt{t_+ -q^2}+ 
\sqrt{t_+-t_0}) \nonumber\\  &\times&\!\!
 \frac{t_+-q^2}{(t_+-t_0)^{1/4}} \times (\sqrt{t_+ -q^2}+
 \sqrt{t_+})^{-5}  \nonumber \\ &\times&\!\!
 (\sqrt{t_+ -q^2}+ \sqrt{t_+-t_-})^{3/2},     
\eea
where $\chi_{1^-}(0)$ is the derivative of the transverse component of
the polarization function $\Pi_{\mu\nu}(q)$ at the Euclidean momentum
$Q^2=-q^2=0$ \cite{Lellouch:1995yv}.  Because of the large value of the
$b$-quark mass, this quantity can be computed by means of perturbative
QCD and the operator product expansion \cite{Generalis:1990id,Lellouch:1995yv}.
On the other hand, the spectral function associated with $\Pi_{\mu\nu}(q)$ is a
sum of positive contributions. Thus, if we assume that it is saturated by
$B\pi$ intermediate states, unitarity and crossing symmetry guarantee
that the coefficients
$a_n(t_0)$ satisfy the inequality
\be\label{unitak}
\sum\limits_{n=0}^\infty a_n^2(t_0) \le 1\,.
\ee
This allows one to calculate bounds on the values of the form factor or
its derivatives at points inside the analyticity domain, in particular
in the physical region. For the $B\to\pi$ form factors, such bounds were
investigated in \cite{Arnesen:2005ez} and \cite{Lellouch:1995yv}.

The expression (\ref{AGRS}) was also adopted as a parametrization of
the form factor in \cite{Boyd:1994tt,Boyd:1995sq,Ball:2006jz}. In this
case the expansion is truncated at a finite order. However, as
noticed in \cite{Becher:2005bg}, the form factor increases then like
$f_+(q^2)\sim (q^2)^{1/4}$ at large $|q^2|$, in contradiction with
perturbative QCD scaling. This behavior follows from the expression
(\ref{phi}) for the outer  function, taking into account that all
 the other factors in \eq{AGRS} are finite at  $|q^2|\to\infty$.  
Moreover, when the series is truncated, the expression
(\ref{AGRS}) has an unphysical singularity at the $B\pi$ production
threshold $t_+$, produced by the factor $(t_+-q^2)$ in the numerator
of \eq{phi}.  This unphysical singularity may distort the behavior
near the upper end of the physical region, where the form factor is
poorly known.

It should be noted that in the calculation of bounds one uses  the full expansion in \eq{AGRS}, with an infinite
number of terms. Then the series cancels the zeros of the function
$\phi(q^2, t_0)$ at $q^2=t_+$ and at $q^2\to\infty$, restoring the required properties of $f_+(q^2)$.  Actually, it can
be shown that imposing the condition that the series vanishes at
threshold or at infinity does not change the unitarity bounds (this
answers a question raised in \cite{Becher:2005bg} about the possibility of
improving the bounds in this way).

A second type of parametrization, used recently for the $B\to\pi$ form
factors in \cite{Flynn:2006vr,Flynn:2007ii}, is based on the Omn\`es
representation \cite{Omnes:1958hv}, which expresses an analytic function in
terms of its phase along the boundary of the analyticity domain.  If
the phase $\delta(t)$, defined by $f_+(t+i\epsilon)= |f_+(t)| \exp(i
\delta(t))$ for $t\ge t_+$, has a finite limit at infinity, the
Omn\`es representation requires only one subtraction. Taking into
account the pole at $q^2=m_{B^*}^2$ and assuming, as in \cite{Flynn:2006vr},
that the form factor does not have zeros in the complex plane, the
representation reads
\bea\label{Omnes} f_+(q^2) = \!\!&f&\!\!\!\!_+(q_1^2)\,\frac{m_{B^*}^2-q_1^2}{m_{B^*}^2-q^2} 
\exp\left[\frac{q^2-q^2_1}{\pi} \right. \nonumber \\  &\times& \!\! \left. \int\limits_{t_+}^{\infty}\frac{\delta(t)\,\mbox{d} t}{(t-q^2_1) (t-q^2)}\,\right], \eea
where $q_1^2$ is an arbitrary subtraction point.  By Watson's theorem
\cite{Watson:1954uc}, the phase $\delta(t)$ is equal, below the first
inelastic threshold, to the phase of the $P$-wave with $I=1/2$ of the
$\pi B\to\pi B$ elastic scattering.  Since this phase is not known, 
in Refs. \cite{Flynn:2006vr,  Flynn:2007ii}  the contribution of the integral was suppressed by
using a multiply-subtracted dispersion relation. Neglecting altogether
the dispersion integral, the form factor is represented in
\cite{Flynn:2007ii} as
\be\label{Omnesms1} f_+(q^2) = \frac{1}{m_{B^*}^2-q^2}
\prod\limits_{j=1}^n[f_+(q^2_j)(m_{B^*}^2-q^2_j)]^{\alpha_j(q^2)},
\ee
where  
\be\label{alphaj}
\alpha_j(q^2)= \prod\limits_{i=0, i\ne j}^n\frac{q^2-q^2_i}{q^2_j-q^2_i}\,.
\ee
However, it is easy to see that the expression (\ref{Omnesms1})
defines an entire function in the complex $q^2$-plane, with no cut for
$q^2\ge t_+$. So, this parametrization does not have the proper
structure of the physical form factor required by analyticity and
unitarity. Moreover, the expression (\ref{Omnesms1}) exhibits at
$|q^2|\to\infty$ an exponential behavior like $\exp[C (q^2)^{n-1}]$,
where $C$ is a combination of the values $f_+(q^2_j)$.
This anomalous behavior follows from the multiple subtraction of
a dispersion relation that requires in general only one subtraction.
The values $f_+(q^2_j)$ for $j>1$ are not independent: according to
(\ref{Omnes}), they can all be expressed in terms of $f_+(q^2_1)$ and
the dispersion integral. By taking into account these relations in the 
multiply-subtracted dispersion relation, one recovers the original
relation (\ref{Omnes}). However, if the values $f_+(q^2_j)$ are
treated as independent, the form factor behaves as an exponential at
large $q^2$, in contradiction with QCD scaling.

Though the shortcomings discussed in this section formally concern the
behavior of the parametrizations of \eqs{AGRS}{Omnesms1} outside the
semileptonic domain, it is important to construct a representation of the
form factor which has the correct analyticity properties in the whole
complex plane. Besides the obvious argument that a parametrization
that does not satisfy these properties cannot describe the form
factor correctly, the introduction of unpysical singularities,
sometimes close to the semileptonic domain, can distort the form
factor in that region. Given the levels of precision currently reached
and expected in the study of exclusive $B\to\pi$ decays, such
distortions are unacceptable.

\section{A new parametrization for $f_+(q^2)$}\label{sec:new}
We start by the remark that the product $ (1-q^2/ m_{B^*}^2) f_+(q^2)$
is analytic in the complex $q^2$-plane cut along the real axis for
$q^2\ge t_+$ and is finite for  $q^2\to\infty$, due to the scaling
behavior $f_+(q^2)\sim 1/q^2$.  An expansion of the product that
converges in the whole complex plane is obtained in terms of a
variable that conformally maps the cut $q^2$ plane onto a disk
\cite{Ciulli:1961zm,Cutkosky:1969iv}.  The variable $z=z(q^2, t_0)$
defined in \eq{z} performs precisely this  mapping. Thus, we
propose the simple parametrization
\be\label{new}
f_+(q^2)= \frac{1}{1-q^2/ m_{B^*}^2}\,\sum\limits_{k= 0}^{K}
b_k(t_0)\, z^k\,.
\ee 
The polynomial in powers of $z$ displays the branch point at
$q^2=t_+$ and is finite in the disk $|z|\le 1$, {\em i.e.} in the
whole $q^2$-plane. This ensures the correct analytic structure in the complex plane and the proper scaling, $f_+(q^2)\sim
1/q^2$ at large $q^2$.  

As mentioned in the Introduction, $f_+(q^2)$ must satisfy also the
condition $\mbox{Im}\, f_+(q^2)\sim (q^2-t_+)^{3/2}$ near $t_+$. Also,
analyticity implies that near threshold $\mbox{Re}\, f_+(q^2)\sim
a_++b_+ (q^2-t_+)+\ldots$, where $a_+$ and $b_+$ are constants. We
recall that from the definition (\ref{z}) of the variable
$z=z(q^2,t_0)$ it follows that the threshold $t_+$ is mapped onto the
point $z=-1$, and $(z+1) \sim {\rm const.} \times (q^2-t_+)^{1/2}$
near $z=-1$. Then, it is easy to see that $f_+$ must satisfy the
condition
\be\label{threshold}
\left[\frac{{\rm d} f_+}{{\rm d} z}\right]_{z=-1}=0,
\ee 
which, written in terms of the coefficients $b_k$ appearing in (\ref{new}), takes the simple form
\be\label{bkthreshold}
\sum_{k=1}^{K}(-1)^{k+1}\, k \, b_k(t_0) =0.
\ee 
By  inserting in \eq{new} the solution of (\ref{bkthreshold}), written as
\be\label{bK}
b_{K}=-\frac{(-1)^{K}}{K}\sum\limits_{k=0}^{K-1} (-1)^k k b_k,
\ee
 we arrive at the expression
\be\label{implem}
f_+(q^2)= \frac{1}  {1-q^2/ m_{B^*}^2}\,  \sum\limits_{k=0}^{K-1} b_k \left[z^k -(-1)^{k-K}\frac{ k}{ K}\, z^{K}\right],
\ee 
where $z= z(q^2,t_0)$. This is the parametrization that we investigate in the present work.

As concerns the conformal mapping, {\em i.e.} the parameter $t_0$ in (\ref{z}), it was remarked in \cite{Boyd:1994tt,Boyd:1995sq} that for
$t_0=t_{opt}$ with
\be\label{t0}t_{opt}\equiv
(m_B+m_\pi)(\sqrt{m_B}-\sqrt{m_\pi})^2=20.062 \,\mbox{GeV}^2\ , \ee
the semileptonic domain is mapped onto the symmetric interval $|z|\le
0.279$ in the $z$-plane. As we shall discuss below, this choice  minimizes the
maximum truncation error in the semileptonic domain.
 An additional argument for this choice is provided by the general theory of
 the representation of data distributed along an interval: as shown in
\cite{Cutkosky:1969iv}, in this case the optimal expansion of the function is obtained by using a complete set of orthogonal polynomials  of the
variable that  maps the original complex cut plane onto an ellipse, such
that the cut becomes the boundary and the physical range is mapped onto the interval situated between
the focal points. We checked that  the optimal
ellipse given in \cite{Cutkosky:1969iv} is in our case very close to the circle $|z(q^2,t_{opt})|=1$.

Other choices of $t_0$ are useful if one is interested in having a more accurate description in a specific energy range. Two values, $t_0=0$ and  $t_0=t_-$ were investigated  in \cite{Ball:2006jz}.  In our analysis we shall adopt the choice $t_0=t_{opt}$, with $t_{opt}$ given in (\ref{t0}). 

\section{Unitarity constraint}\label{sec:unitarity}
The unitarity condition (\ref{unitak}) can  also be expressed in
terms of the coefficients $b_k(t_0)$.  By comparing the
representations in \eqs{AGRS}{new} we have
\be\label{equality} \sum\limits_{n=0}^\infty a_n(t_0) z^n = \Psi(z)
\sum\limits_{k=0}^{K} b_k(t_0) z^k \,, \ee
where $\Psi(z)$ is a known function
\be\label{Psi} \Psi(z)= \frac{ m_{B^*}^2}{4 (t_+-t_0)}\, \Phi(z)\,
\frac{(1-z)^2 (1-z_*)^2}{(1-z z_*)^2}.  \ee
We denote by $z_*=z(m_{B^*}^2, t_0)$ the position of the pole in the
variable $z$, and $\Phi(z)\equiv \phi(q^2(z), t_0)$ is the outer
function expressed in terms of $z$, by using the inverse of \eq{z}
\be\label{tz}
q^2(z)= t_+ -(t_+-t_0)\, \left(\frac{1+ z}{1- z} \right)^2\,.
\ee
The function $\Psi(z)$, which depends also on the parameter $t_0$, is
analytic in $|z|<1$. Thus, we can expand it around $z=0$ as
\be\label{phiexpan}
\Psi(z)=\sum\limits_{k=0}^\infty \eta_k(t_0) z^k.
\ee
Inserting this expansion in (\ref{equality}), we obtain
\be\label{atob} a_n(t_0)=\sum\limits_{k=0}^{\min[K, n]}
\eta_{n-k}(t_0) b_k(t_0), \quad\quad n\ge 0.  \ee
Then the inequality (\ref{unitak}), expressed in terms of the
coefficients $b_j(t_0)$, reads as
\be\label{unitbn}
\sum\limits\limits_{j,k=0}^{K} B_{jk}(t_0) b_j(t_0) b_k(t_0) \le 1,
\ee
where 
\be\label{Bjk}
 B_{jk}(t_0) =\sum\limits_{n=0}^\infty \eta_{n}(t_0)\eta_{n+|j-k|}(t_0).
\ee
From \eqs{phi}{Psi} it follows that the function $\Psi(z)$ is
bounded in the closed disk $|z|\le 1$, so its Taylor coefficients
$\eta_j$ are rapidly decreasing. Therefore, the coefficients
$B_{jk}(t_0)$ can be computed by performing in (\ref{Bjk}) the
summation upon $n$ up to a finite order, about 100 in practice.

As discussed in \cite{Becher:2005bg}, the leading contributions to the
sum over the coefficients $a_n^2$, which appears in the unitarity
condition (\ref{unitak}), are of order $(\Lambda/m_b)^3$ in the heavy
$b$-quark expansion. Thus, we expect that the 1 appearing on the 
right-hand sides of the constraints of Eqs.~(\ref{unitak}) and
(\ref{unitbn}) is a significant overestimate, the real bound being
more realistically on the order of a few per mil. Were we to consider
these more stringent bounds in the sequel, we would be able to reduce
the number of terms kept in the series expansion of the form factor.
This would significantly reduce the systematic uncertainty that we
encounter when using this expansion to extrapolate the form factor to
regions where it is not constrained by the input that we use. However,
in the absence of a more precise quantitative argument for strengthening
the bound, we choose to keep Eq.~(\ref{unitbn}) as it stands. In any
event, if and when such an argument is found, the procedures explained
in the following sections can be carried over as is, simply replacing
the right-hand side of the inequality (\ref{unitbn}) by the relevant
smaller number.

For the numerical evaluation of the coefficients $B_{jk}$ we need the
value of $\chi_{1^-}(0)$ entering the outer function (\ref{phi}).
Perturbative QCD and the operator product expansion give
~\cite{Generalis:1990id,Lellouch:1995yv}
\be\label{chi} \chi_{1^-}(0)=\frac{3 [1+1.14\, \alpha_s(\bar m_b)]}{32 \pi^2
  m_b^2}-\frac{\bar{m}_b\langle\bar{u}u\rangle}{m_b^6}-\frac{\langle\alpha_s
  G^2\rangle}{12 \pi m_b^6}\ , \ee
where  $m_b= 4.9\, \mbox{GeV}$ is  the pole mass and $\bar{m}_b$ the $\overline{\mbox{MS}}$ mass, with
 $\bar{m}_b(2\,\mbox{GeV})\approx 4.98$ GeV, obtained from
 $\bar{m}_b(\bar{m}_b )\approx 4.2$ GeV \cite{PDG:2008} and the four-loop running in the $\overline{\mbox{MS}}$ scheme \cite{Chetyrkin:2000}. We took $\alpha_s(\bar{m}_b)=0.22$ \cite{PDG:2008}. The gluon condensate has the standard value   $\langle\alpha_s G^2\rangle = 0.038\,\mbox{GeV}^4$ given in \cite{SVZ},
 while for the quark condensate we used the two-flavor value $\langle \bar{u}u\rangle \approx -(278\,\,  \mbox{MeV})^3$  in the  $\overline{\mbox{MS}}$ scheme at scale 2 GeV \cite{Bernard}.  From the above values we derive  $\bar{m}_b\, \langle\bar{u}u\rangle \approx -0.107\ \mbox{GeV}^4$ at scale 2 GeV, and we adopt this value also at scale $m_b$, since the scale dependent corrections to the product are negligible. 
By inserting  in (\ref{chi}) the above central values we obtain  $\chi_{1^-}(0)\approx 5.01  \times 10^{-4}$. 
 For illustration, we
give in \tab{tab:Bijt0} the coefficients $B_{jk}(t_0)$ calculated with this value of  $\chi_{1^-}(0)$  for $K=5$ and several values of $t_0$. 

We mention also that when $K\to \infty$, the expansion
in \eq{new} is convergent in the whole disk $|z|<1$, {\em i.e.} in the
whole $q^2$-plane cut along the real axis for $q^2\ge t_+$. Moreover, the
unitarity condition (\ref{unitbn}) can be used to derive an explicit upper bound on the truncation error. We present the derivation of this bound  in the Appendix.

\begin{table}[h]\caption{\label{tab:Bijt0} The matrix elements, $B_{jk}(t_0)$,  which 
enter the unitarity bound (\ref{unitbn}) for $K=5$ and several values of $t_0$. The remaining 
  coefficients are obtained from the relations
  $B_{j(j+k)}=B_{0k}$ and the symmetry property
  $B_{jk}=B_{kj}$,  obvious from \eq{Bjk}. }
\begin{center}\begin{tabular}{lccccccc}
\hline\hline
$t_0$(${\rm GeV^2}$) &$B_{00}$&$B_{01}$&$B_{02}$  &$B_{04}$&$B_{04}$ &$B_{05}$ \\\hline
0& 0.0197&-0.0049& -0.0108 & 0.0057  &0.0006 &-0.0005\\
$t_{opt}$ & 0.0197&0.0042& -0.0109& -0.0059 &-0.0002& 0.0012 \\
$t_-$&0.0197&0.0118  &-0.0015 & -0.0078 &-0.0077& -0.0053\\
\hline\hline
\end{tabular}\end{center}

\end{table}

\section{ Theoretical and experimental input}\label{sec:input}
At low $q^2$, the form factor is calculated in the framework of LCSR
\cite{Khodjamirian:1997ub,Ball:2004rg,Duplancic:2008ix}. We use
$f_\mathrm{LCSR}\equiv f_+(0)=0.26$, with the uncertainty $\delta
f_\mathrm{LCSR}=0.03$ \cite{Ball:2006jz}.  Lattice calculations
provide the value of the form factor at eight additional $q^2$-points:
three are taken from FNAL-MILC \cite{Okamoto:2004xg,Mackenzie:2005wu}
and five from HPQCD, updated in \cite{Dalgic:2006dt}. As in
\cite{Flynn:2006vr,Flynn:2007ii}, we take the three FNAL-MILC results
from \cite{Arnesen:2005ez}.~\footnote{As we were finalizing this work,
  Fermilab and MILC presented, in~\cite{Bailey:2008wp}, a substantial
  update of their lattice calculation of the form factor
  $f_+(q^2)$. Since their new results agree within errors with those
  of \cite{Okamoto:2004xg,Mackenzie:2005wu} and since our goal here is
  to illustrate the workings of our new parametrization, we have
  chosen not to update our analysis. Instead, we encourage the authors
  of \cite{Bailey:2008wp} to perform their analysis with our improved
  parametrization.}

The available experimental data consist in the partial branching
fractions over bins in $q^2$. We use 10 data from the tagged analyses
(4 bins from CLEO \cite{Adam:2007}, which replace the older data
\cite{Athar:2003yg}, 3 from Belle \cite{Hokuue:2006nr}, and 3 from
BaBar \cite{Aubert:2006}), and 12 bins from the untagged BaBar
analysis \cite{Aubert:2007px}, where the full covariance matrix is
available.  The total number of data points from theory and experiment
is 31.

It is convenient to define the global $\chi^2$ 
\be\label{chi2}
\chi^2(b_k, |V_{ub}|)=\chi^2_{th}+\chi^2_{exp}\,,
\ee
where
\bea\label{chi2th} && \hspace{-0.6cm}\chi^2_{th}= \sum\limits_{j,k=1}^8 [f_j^{in} -
f_+(q^2_j)]C^{-1}_{jk}[f_k^{in} - f_+(q^2_k)] \nonumber\\ && ~~~~~~~ +
(f_+(0)\nonumber -f_\mathrm{LCSR})^2/(\delta f_\mathrm{LCSR})^2
\,,\nonumber\\
&&\hspace{-0.6cm}\chi^2_{exp} = \sum\limits_{j,k=1}^{22} [{\mathcal B}_j^{in} -
{\mathcal B}_j(f_+)]C^{-1}_{{\mathcal B}\,jk}[{\mathcal B}_k^{in} -{\mathcal B}_k(f_+)] \,.  \eea
In the above relations, the $f_j^{in}$ denote the values of the form
factor calculated on the lattice at the points $q^2_j$; ${\mathcal B}_j^{in}$ are
the experimental partial branching fractions and ${\mathcal B}_j(f_+) $ the
values calculated by integrating \eq{spectrum} over the bins $[q^2_j,
q^2_{j+1}]$,  with a given parametrization for the form factor
$f_+(q^2)$. To convert to a rate, we use the $B^0$ lifetime, $\tau_B^0
=1/\Gamma_{\rm tot}= (1.527 \pm 0.008)\times 10^{-12} s$
\cite{Barberio:2007cr}.

 The covariance matrices $C$ and $C_B$ are
written formally: in practice they are block diagonal, with independent
blocks for each independent set of experimental data or lattice
results. Unfortunately, the covariance matrices are not provided for
the lattice calculations. Thus, we make here a set of reasonable
assumptions on the possible correlations based on the information
provided in the papers and on our experience with such
calculations. The lattice results of
\cite{Okamoto:2004xg,Mackenzie:2005wu} and \cite{Dalgic:2006dt} are
obtained using different discretizations for the heavy quark,  but
on subsets of the MILC, $N_f=2+1$, gauge configurations which have
significant overlap. Thus, in addition to assuming that the
statistical errors on $f_+(q^2)$ at different $q^2$ within each
calculation have a 50\% correlation,\footnote{Points at different
  $q^2$ within a given simulation are obtained on the same statistical
  ensemble with very similar methods and are thus expected to be
  strongly correlated. A glance at \fig{fig:fplus4vsqsq}, in which the
  lattice results are plotted, should convince the reader that such
  correlations are present.} we assume that there is a 25\%
correlation between the errors in the two calculations. Such
correlations in the statistical errors have been assumed to be
negligible in previous work
\cite{Arnesen:2005ez,Flynn:2006vr,Flynn:2007qd,Flynn:2007ii}. Regarding the
systematic errors, since the heavy-quark discretizations and methods
used are different, we assume negligible correlations in the
systematic errors between the two calculations, but given their
nature, assume 100\% correlations within each simulation. Though these
assumptions cannot replace covariance matrices determined by the
lattice collaborations themselves, we believe that they are reasonable
and will not lead to underestimated errors. We have verified that
without these correlations, for instance, the results for $f_+$ versus
$q^2$ quoted below would have fit errors reduced by up to 30\%.

\section{Results of the fits}\label{sec:fits}

We performed a combined fit of the above input by minimizing  $\chi^2$ defined in (\ref{chi2}), using the  representation of $f_+(q^2)$ given in (\ref{implem}), with $z=z(q^2, t_{opt})$. The free parameters are
$|V_{ub}|$ and the real coefficients $b_k$, $k\le K-1$  subject to  the unitarity constraint  (\ref{unitbn}), with $b_{K}$ given by the expression  (\ref{bK}). The total number of parameters is $N=K+1$.

According to convex optimization theory \cite{Luen}, the optimum
values $b_k^{(0)}$ and the optimal Lagrange multiplier $\lambda_0$
minimizing the Lagrangian
\be\label{Lagr}
 {\cal L}(b_j,|V_{ub}|)=\chi^2(b_j, |V_{ub}|)  + \lambda
 \left(\sum\limits\limits_{j,k=0}^{K} B_{jk}
 b_j b_k - 1 \right) \,,
\ee
satisfy the alignment condition
\be\label{allign}
 \lambda_0
 \left(\sum\limits\limits_{j,k=0}^{K} B_{jk}
 b_j^{(0)} b_k^{(0)} - 1 \right)=0.
\ee
Therefore, either $\lambda_0=0$ and the optimal parameters $b_k^{(0)}$
of the unconstrained minimum of $ \chi^2(b_k, |V_{ub}|)$ satisfy
automatically the constraint (\ref{unitbn}), or $\lambda_0\ne 0$, when
the optimal parameters saturate the constraint (\ref{unitbn}).

\begin{figure*}[t]\begin{center}
\includegraphics[width=7.3cm]{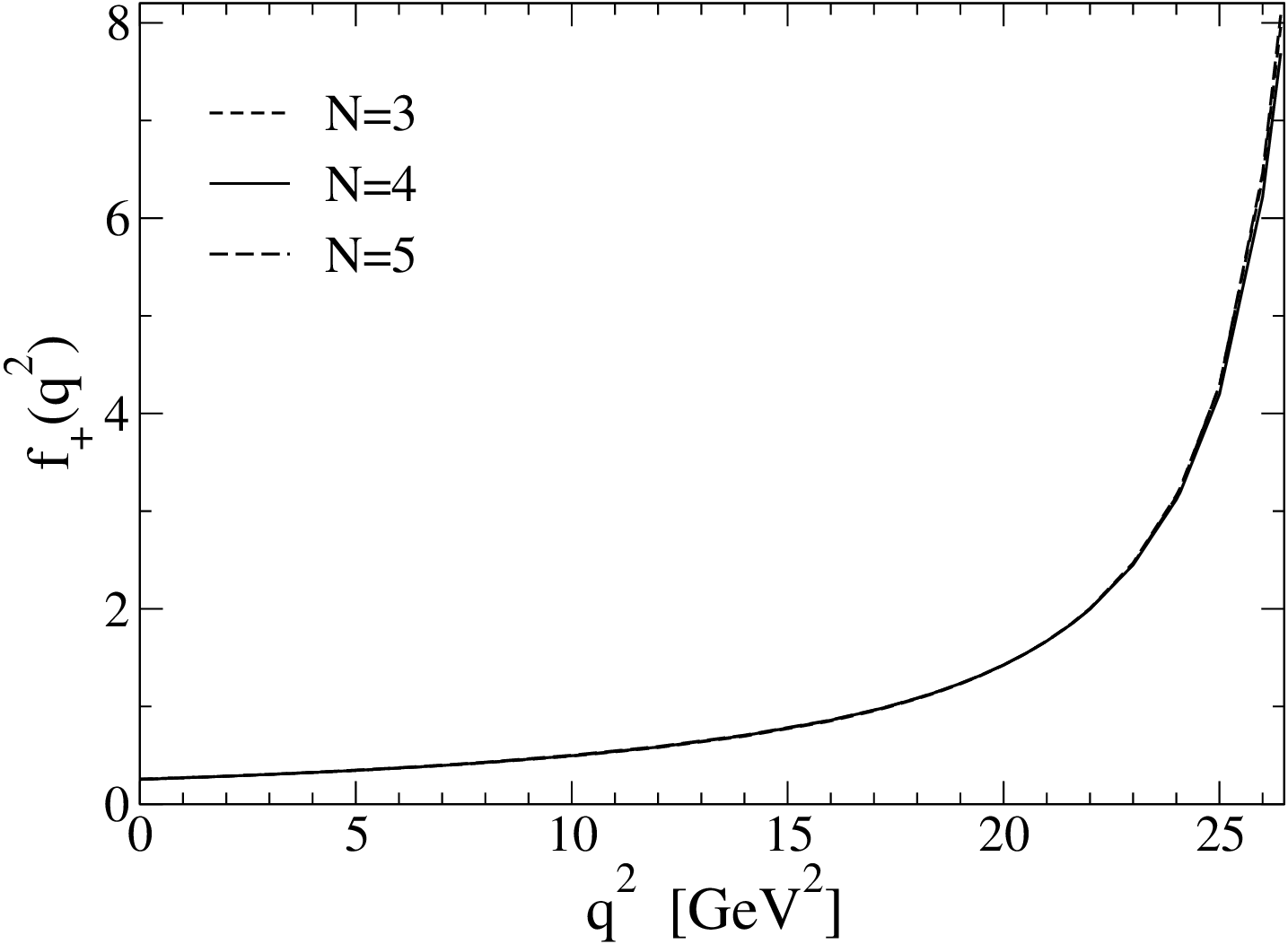}\hspace{0.8cm}
\includegraphics[width=7.3cm]{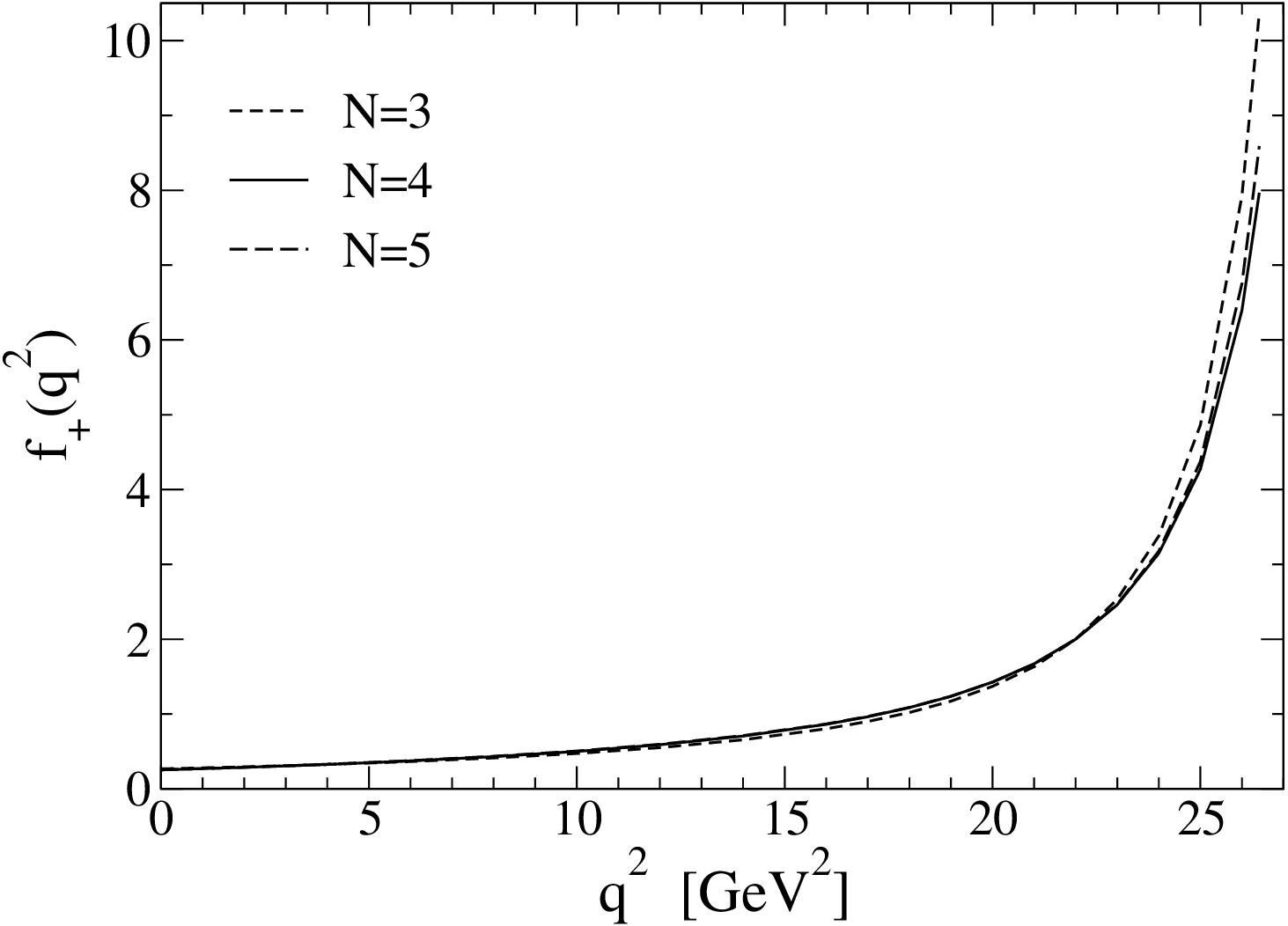}
\caption{\label{fig:newvstrad} Left: the  form factor 
  $f_+(q^2)$ computed with the new representation (\ref{implem}) and the parameters from (\ref{N3})-(\ref{N5}). Right: the form factor calculated  with  the traditional parametrization (\ref{AGRS}) and the parameters from  (\ref{N3t})-(\ref{N5t}).}
\end{center}\vspace{-0.5cm}
\end{figure*}

A nontrivial form factor is obtained for $K\ge 2$, {\em i.e.} a total
number of parameters, $N\ge 3$.  The results of the fits obtained by
increasing $N$ are presented below:

\bea\label{N3}
 N=3; && K=2;\quad  \chi^2= 21.17,\quad \chi^2/dof=0.76,\nonumber\\
&& \hspace{-1.cm} b_0=0.420 \pm 0.031, \quad b_1=-0.514 \pm 0.070,\nonumber\\
&&\quad\quad  |V_{ub}|=(3.58 \pm 0.27)\times 10^{-3}, \eea
\bea \label{N4}
&&N=4; \quad \quad K=3;\quad \quad \chi^2= 21.03, \nonumber \\
&&\chi^2/dof=0.78, \quad\quad b_0=0.421 \pm 0.031, \nonumber\\ 
&&\hspace{-0.5cm} b_1=-0.476 \pm 0.122, \quad \quad b_2=-0.399 \pm 0.381, \nonumber\\
&&\hspace{1cm}|V_{ub}|=(3.57 \pm 0.27)\times 10^{-3}, \eea
\bea\label{N5}
&&\quad \quad N=5; \quad \quad K=4; \quad \quad \chi^2= 21.00, \nonumber \\ 
&&\quad \chi^2/dof=0.81, \quad  b_0=0.421 \pm 0.031,  \\ 
&& b_1=-0.469 \pm 0.129, \quad  b_2=-0.178\,^{+1.358}_{-1.313}\,,\nonumber \\  
&& b_3=-0.825\,^{+4.067}_{-4.042}\,, \quad |V_{ub}|=(3.54 \pm 0.30)\times 10^{-3}. \nonumber\eea
 For simplicity we omitted the upper index ``(0)'' in the notation of the
 optimal parameters. All the errors indicated are statistical. We
 mention that for the fits (\ref{N3}) and (\ref{N4}) the unitarity
 constraint (\ref{unitbn}) is not saturated, while the  slightly asymmetric
 errors on the coefficients $b_2$ and $b_3$ in (\ref{N5}) are produced
 mainly by this constraint.

 The form factor calculated with the central values of the parameters
 from (\ref{N3})-(\ref{N5}) is plotted in the left panel of
 Fig. \ref{fig:newvstrad}.  For comparison we repeat the analysis also
 with the standard parametrization (\ref{AGRS}).  We recall that this
 parametrization has an unphysical singularity at threshold, and we
 cannot impose the threshold condition that we use above. Therefore, for a
 certain $K$ there are $K+1$ parameters $a_k$, constrained by the
 unitarity condition (\ref{unitak}), and the total number of
 parameters is $N=K+2$. The best fits for the lowest values of $N$
 are
\bea\label{N3t}
 N=3; && K=1;  \quad \chi^2=27.68,\quad  \chi^2/dof=0.98,\nonumber\\
&&  \hspace{-1.cm} a_0=0.024 \pm 0.002,\quad  a_1=-0.033 \pm 0.004,\nonumber\\
&& \quad\quad|V_{ub}|=(3.63 \pm 0.28)\times 10^{-3}, \eea
\bea \label{N4t}
&&N=4; \quad \quad  K=2;\quad \quad \chi^2= 21.04, \nonumber \\
&&\chi^2/dof=0.79,\quad\quad  a_0=0.025  \pm 0.002,\nonumber \\
&&\hspace{-0.5cm} a_1= -0.021 \pm 0.007, \quad\quad a_2=-0.067 \pm 0.026, \nonumber\\
&&\hspace{1.cm} |V_{ub}|=(3.54 \pm 0.26)\times 10^{-3}, \eea
\bea\label{N5t} 
&& \quad \quad N=5; \quad K=3; \quad \quad \chi^2= 21.01,\nonumber \\ 
&&\chi^2/dof=0.81,\quad  a_0= 0.025 \pm 0.002,\\
&&a_1= -0.020 \pm 0.008, \quad a_2=-0.054 \pm 0.075, \nonumber \\ 
&& a_3=-0.056 \pm 0.308,  \quad|V_{ub}|=(3.52 \pm 0.29)\times 10^{-3}.\nonumber \eea
In all cases, the unitarity constraint (\ref{unitak}) is not
saturated.  The corresponding form factor is plotted in the right
panel of Fig.\ref{fig:newvstrad}.  For large values of $q^2$ the
results indicate a more pronounced variation with $N$ than that of
the curves in the left-hand panel of the figure.

\section{Systematic error}\label{sec:syst}
In the present framework, the systematic error on the values of the
form factor must account for the effect of truncating the expansion
(\ref{new}) at a finite order $K$. As shown in the appendix, the
unitarity constraint (\ref{unitbn}) can be exploited to derive an
upper bound on the truncation error. However, this estimate is too
conservative for low values of $K$. A more realistic prescription is
given by the magnitude of the next term in the expansion, allowed by
the unitarity constraint.  Denote by $b^{max}_{K+1}$ the maximum value
of the modulus $|b_{K+1}|$, allowed by the condition (\ref{unitbn}),
for fixed values of $b_k$, $k\le K$, given by the fit. We note that,
although the inequality (\ref{unitbn}) may be saturated by the latter
values, as happens with the values in Eq. (\ref{N5}), a nonzero value
for $b^{max}_{K+1}$ is obtained, since the convex condition
(\ref{unitbn}) is not a sum of squares.

According to the above discussion, we  adopt  as a realistic prescription for the systematic error on the form factor the quantity
\be\label{ffsyst} 
\delta f_+(q^2)_{syst}= \frac{b^{max}_{K+1}\, |z^{K+1}|}{1-q^2/m_{B^*}^2}, 
\ee
where $z=z(q^2, t_{opt})$.  Using  the optimal $b_k$ from (\ref{N3})-(\ref{N5}), $b_K$ from (\ref{bK}) and the values of $B_{jk}$ for $t_0=t_{opt}$ given in Table \ref{tab:Bijt0}, we obtain
\be\label{bmax}
b^{max}_{3}= 6.97,   \quad \quad b^{max}_{4}= 6.74,   \quad \quad  b^{max}_{5}= 6.51.
\ee 
With these values,  the numerator of (\ref{ffsyst}) calculated at the limits of the physical region, $|z_{max}|=0.279$, is  a fraction of 36\%, 9.7\% and 2.6\%  from the corresponding first coefficient $b_0$ given in (\ref{N3})-(\ref{N5}), for $N=3$, $N=4$ and $N=5$, respectively. 
  For illustration we give also the values of the form factor and errors at the highest point $t_{-}$, where the systematic error defined in (\ref{ffsyst}) has the largest value
\bea\label{statsyst} 
&&\hspace{-0.6cm} f_+(t_{-})= 7.96 \pm 0.70\, {\rm (stat)} \pm 2.23 \,{\rm (syst)},\quad\quad N=3,\nonumber\\
&&\hspace{-0.6cm}f_+(t_{-})= 7.69 \pm 1.00\, {\rm (stat)} \pm 0.60\, {\rm (syst)}, \quad\quad N=4,\nonumber\\
&&\hspace{-0.6cm}f_+(t_{-})= 8.08 \,\,^{+2.53 }_{-2.45 }\,\,\,\,\, {\rm (stat)} \pm 0.16\, {\rm (syst)}, ~~ N=5.  
\eea
For $N=5$ the systematic error is very small. Actually, it is negligibly small compared to the statistical error  along the whole physical region. By going up to  $N=5$, we can neglect the systematic error  altogether for the determination of $V_{ub}$ and for the form factor in the physical region. We shall adopt this choice as our optimal parametrization.

\begin{table*}[floatfix]\caption{\label{tab:fplusqsq1} The form factor at a variety of $q^2$
  values in the semileptonic domain, as obtained with the new expression
  of \eq{optim} and the parameters given in \eq{bestfit}. The errors
   are obtained as described in the text.}
 \begin{center}\begin{tabular}{llll}\hline\hline
  $q^2\,(\mbox{GeV}^2)\hspace{1cm}$ &  $~~~f_+(q^2) \hspace{2cm} $ & $
q^2\,(\mbox{GeV}^2)\hspace{1cm}$ &  $~~~f_+(q^2)$  \\*[0.1cm]
\hline\\*[-0.4cm]
  0. & $0.254\,^{+ 0.023}_{- 0.022}$  & 18. &  $1.086 \,^{+0.087 }_{-0.086 }$  \\
  2.& $0.287\,^{+ 0.024}_{- 0.024}$  & 19.& $ 1.237\,^{+0.098 }_{-0.097 }$  \\
  4.& $0.326 \,^{+ 0.029}_{-0.028}$  & 20.&  $1.425 \,^{+0.114 }_{- 0.113}$  \\
  6.& $0.373\,^{+0.036}_{-0.034}$  & 21.&  $1.670 \,^{+0.135 }_{-0.134 }$  \\
  8.&  $0.430 \,^{+0.043 }_{-0.041 }$   &  22.& $1.998 \,^{+0.166 }_{- 0.164}$  \\
  10.&  $0.501 \,^{+0.050 }_{-0.048 }$  & 23.&  $ 2.458\,^{+ 0.218}_{-0.217 }$ \\
  12.& $ 0.590\,^{+0.057 }_{-0.055 }$   & 24.&  $3.148 \,^{+0.339 }_{-0.335 }$ \\
  14.&  $ 0.707\,^{+ 0.064}_{-0.062 }$ & 25.&  $4.283 \,^{+0.666 }_{-0.652 }$ \\
  16.&  $0.864 \,^{+0.072 }_{-0.071 }$  & 26.&  $6.461 \,^{+1.629 }_{-1.578 }$ \\
  17.& $0.965 \,^{+0.078 }_{-0.077 }$  & 26.42&  $8.080 \,^{+2.533 }_{-2.445 }$ \\*[0.05cm]
\hline\hline
\end{tabular}
\end{center}\end{table*}

\section{Best parametrization in the physical range}\label{sec:best}

As discussed above,  we adopt  the expansion (\ref{implem}) for $K=4$, which writes as
\be\label{optim}
f_+(q^2)= \frac{1}  {1-q^2/ m_{B^*}^2}\,  \sum\limits_{k=0}^{3} b_k \left[z^k -(-1)^{k}\,\frac{k}{4}\, z^{4}\right].
\ee 
The best parameters and their statistical errors, already given in (\ref{N5}), are
\bea\label{bestfit}
&& b_0=0.42 \pm 0.03,\quad\quad b_1=-0.47\pm 0.13,\nonumber\\ 
&& b_2=-0.18\,\pm 1.34, \quad  b_3=-0.83 \pm 4.05.
\eea
We mention that the parameters are correlated, the correlations being highly nongaussian because of the unitarity constraint.  


As shown in (\ref{N5}), the fit gives $\chi^2= 21.00$ and
$\chi^2/dof=0.81$. For completeness we list below the separate
contributions to $\chi^2_{th}$ and $\chi^2_{exp}$ of the various data
sets, compared with the number $n$ of points:
\bea\label{chifin} &&\chi^2_{\rm LCSR} = 0.04 \quad\quad \quad(n=1) \nonumber\\
 &&\chi^2_{\rm FNAL-MILC\,\&\, HPQCD} = 5.13 \quad\quad (n=3+5) \nonumber\\ 
&&\chi^2_{\rm Belle}=0.004\quad\quad\quad  (n=3) \nonumber\\ 
 &&\chi^2_{\rm CLEO}=2.81 \quad\quad\quad (n=4)  \\
&&\chi^2_{\rm BaBar-t}=4.32 \quad\quad  (n=3)\nonumber\\ 
&& \chi^2_{\rm BaBar-u}=8.71\quad\quad (n=12). \nonumber \eea
 The description of all the sets is very good, except for the BaBar  tagged (t) data \cite{Aubert:2006}, where  $\chi^2$ is larger than the number of points.

\begin{figure*}[t]\begin{center}\vspace{0.1cm}
\includegraphics[width=7.3cm]{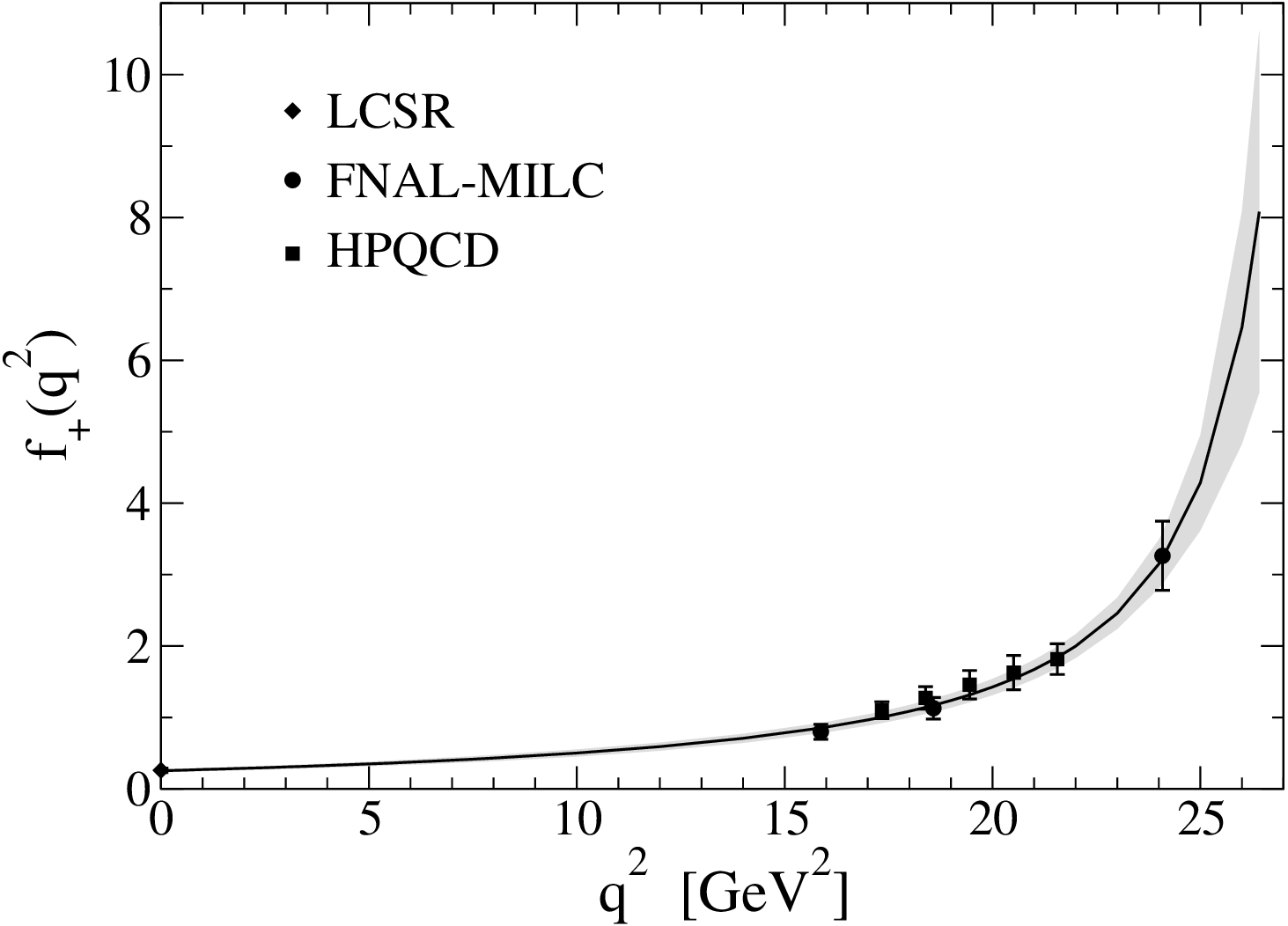}\hspace{0.6cm}\includegraphics[width=7.3cm]{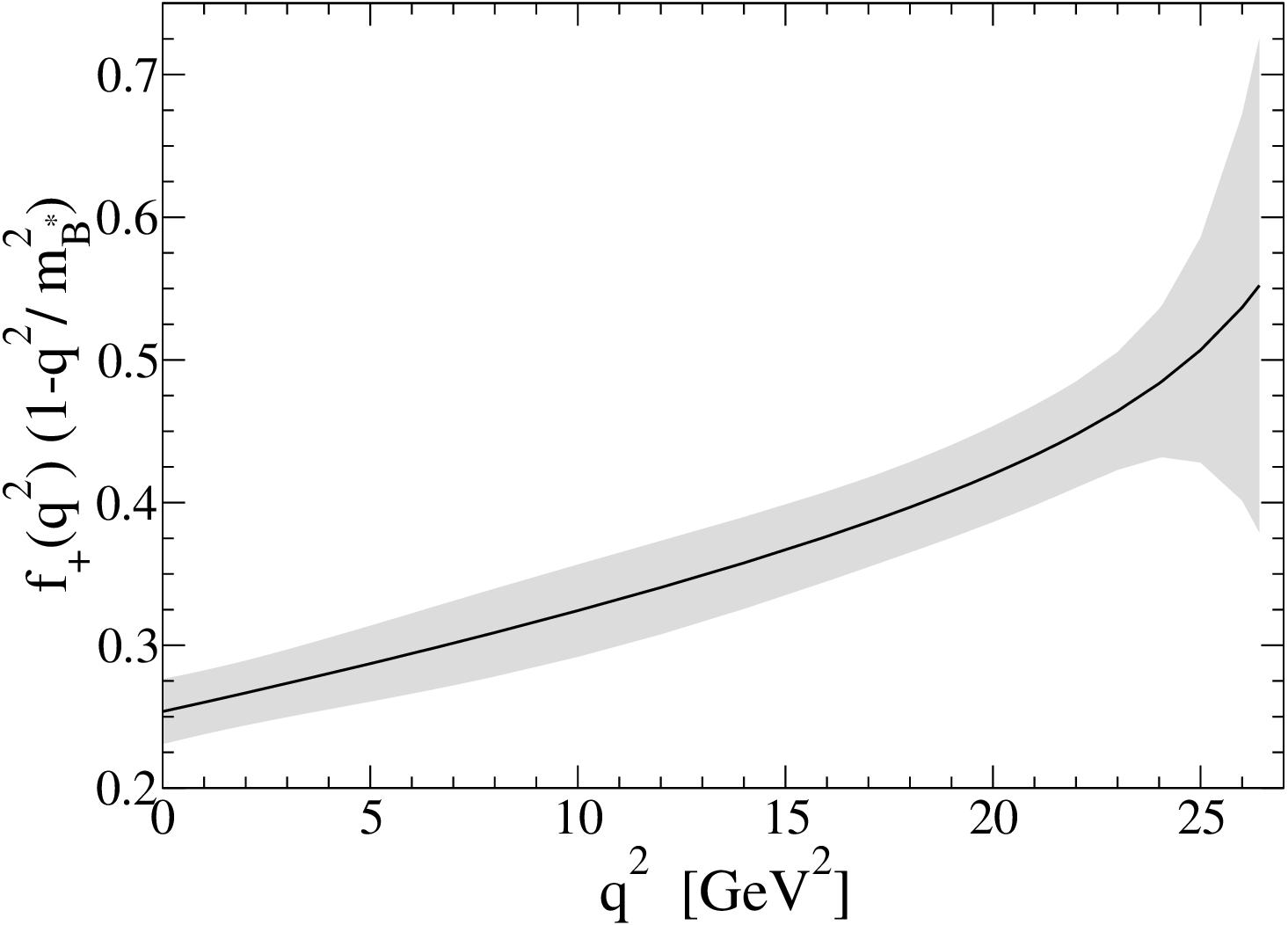}
\caption{\label{fig:fplus4vsqsq} Left: the form factor $f_+(q^2)$
  given by \eqs{optim}{bestfit}. The error band is given  by the statistical uncertainties.  As explained 
in the text, systematic errors coming from the parametrization are 
negligible in the semileptonic domain in our approach.  The theoretical LCSR result from
  \cite{Duplancic:2008ix} and the lattice results from
   \cite{Okamoto:2004xg,Dalgic:2006dt} are also shown. Right: the
  numerator in (\ref{optim}) for the optimal parametrization. }
\end{center}\vspace{-0.5cm}
\end{figure*}
\begin{figure*}[t]\begin{center}\vspace{0.cm}
\includegraphics[width=7.3cm]{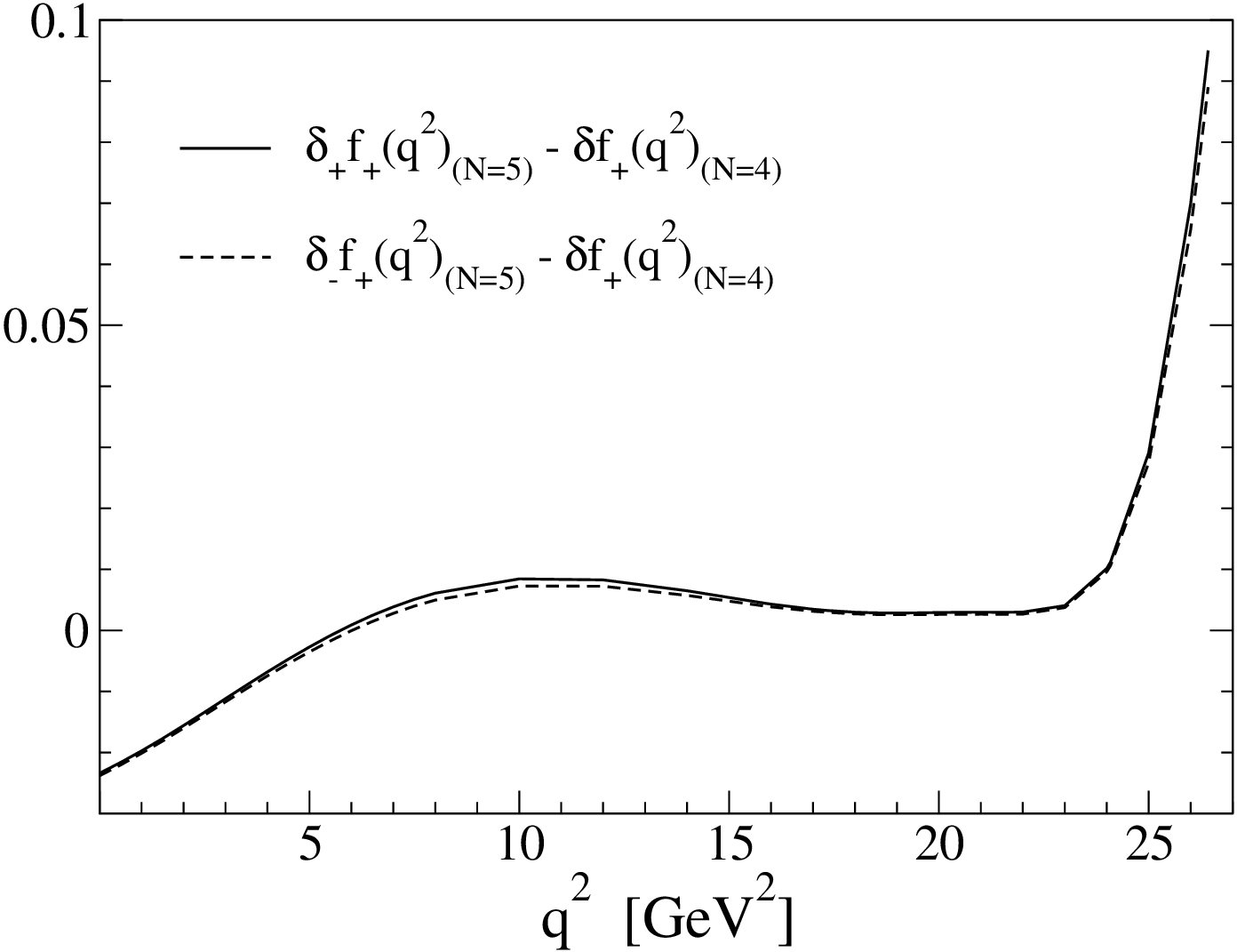}
\caption{\label{fig:dif43} Change in the error with the number of parameters, $N$, in the physical 
region: the solid (dashed) line shows the difference between the $N = 5$ 
and $N = 4$ plus (minus) errors on the product $f_+(q^2)(1- q^2/m^2_{B^*})$. }
\end{center}\vspace{-0.cm}
\end{figure*}

\begin{table*}[htb]\caption{\label{tab:fplusqsq2} Comparison of the form factor
  $f_+(q^2)$ obtained at a variety of recoils in the lattice
  computations of FNAL-MILC \cite{Okamoto:2004xg,Mackenzie:2005wu}
  (rows 1-3) and HPQCD \cite{Dalgic:2006dt} (rows 4-8), with the
  results of the combined fit in \eq{bestfit}, to the parametrization
  given in \eq{optim}. In the lattice results, the first error is the
  combined statistical and chiral extrapolation error, while the
  second is an 11\% systematic for FNAL-MILC
   \cite{Okamoto:2004xg,Mackenzie:2005wu} and a 9.5\% error for
  HPQCD \cite{Dalgic:2006dt}. } \begin{center}
\begin{tabular}{lll}\hline\hline
$q^2\,(\mbox{GeV}^2)\hspace{1cm}$ &~~~~~~~ $f_+(q^2)_{\rm lattice}\hspace{3cm}$ & ~~$f_+(q^2)_{\rm fit}$  \\\hline\\*[-0.4cm]
15.87 &  0.799  $\pm$ 0.058 $\pm$ 0.088  &  $0.852 \,^{+0.071 }_{- 0.070 }$ \\
18.58 &  1.128  $\pm $ 0.086 $\pm$ 0.124 &$ 1.169 \,^{+ 0.093 }_{- 0.092 }$  \\
24.09 & 3.263  $\pm $ 0.324 $\pm$ 0.359  & $ 3.227 \,^{+ 0.356 }_{- 0.352}$ \\
17.34 &  1.101  $\pm $ 0.053 $\pm$ 0.105   & $ 1.003 \,^{+0.081  }_{-0.080}$\\
18.39 & 1.273 $\pm $ 0.099 $\pm$ 0.121  &  $1.141  \,^{+0.091 }_{- 0.090 }$ \\
19.45 & 1.458 $\pm $  0.142 $\pm$ 0.139  & $ 1.316 \,^{+ 0.105 }_{- 0.104 }$ \\
20.51 & 1.627  $\pm $ 0.185  $\pm$ 0.155  & $ 1.542 \,^{+ 0.124 }_{-0.123}$ \\
21.56 &  1.816 $\pm $0.126  $\pm$ 0.173  & $1.841  \,^{+0.151  }_{-0.149  }$ \\*[0.05cm]
\hline\hline
\end{tabular}
 \end{center}
\end{table*}

The form factor calculated using the expression (\ref{optim}) and the
parameters from \eq{bestfit} is shown in \fig{fig:fplus4vsqsq}, where
in the right panel we plot the $z$ polynomial in the numerator of
\eq{optim}. The error bands represent the purely statistical error. We
emphasize that we do not use the linear approximation in the error
propagation, but apply the standard $\Delta \chi^2$ analysis, {\em
  i.e.} by finding the range of variation of a given parameter
corresponding to a change in $\chi^2$, minimized over all other
parameters, by one unit.  The unitarity constraint plays a nontrivial
role, being responsible for the asymmetric errors, especially near the
right end of the semileptonic range.  For completeness, the values of
the form factor are given in \tab{tab:fplusqsq1} for a sample of $q^2$
in the semileptonic domain.  In \tab{tab:fplusqsq2} we compare the
results of our combined fit with the lattice results used as input.

As seen from the values given in (\ref{statsyst}), the gradual
reduction of the systematic error with the increase of $N$ is balanced
by the increase of the statistical error. It is of interest to compare
the total error on the values of the form factor for $N=5$ and
$N=4$. In Fig.~\ref{fig:dif43} we plot the difference between these
two errors as a function of $q^2$. For $N=5$ the error is purely
statistical, for $N=4$ it is calculated by adding quadratically the
statistical and the systematic errors, the later one obtained from
(\ref{ffsyst}).  As we already noted, the error in the $N=5$ case is
slightly not symmetric due to the unitarity constraint, therefore we present
separately the difference between the ``plus'' and "minus" $N=5$ errors
and the $N=4$ one.  Figure \ref{fig:dif43} shows that the difference
between the errors is practically zero for most of the energy range,
including the energies where lattice input is available. At low values
of $q^2$, in particular  at $q^2=0$, the total error decreases when
passing from $N=4$ to $N=5$. On the other hand, at high values of
$q^2$ the error for $N=5$ is larger than the $N=4$
error. One may ask whether it is not
preferable to take as a best prediction the parametrization with
$N=4$. In our opinion, this is not the case: the advantage of our
prescription is that the systematic errors are negligible along the
whole physical region. Thus, we avoid any bias related to the specific
form of the truncation error for the determination of the form factor
and of $V_{ub}$. Our results show that a representation of the form
factor having a small uncertainty over the whole physical region,
including its upper end, is not possible with the present input
information.

 From the above comment we expect even larger errors if the expression (\ref{optim}) is used to calculate the form factor outside the physical region.  In particular, we
consider  the residue 
of $f_+(q^2)$ at the pole $q^2=m_{B^*}^2$, defined as
\be\label{residue}
r_+=\lim_{q^2 \to m_{B^*}^2} (1-q^2/ m_{B^*}^2)\, f_+(q^2).
\ee
The systematic error, calculated using the prescription (\ref{ffsyst}), is no longer negligible at the position $z_*= -0.504$ of the pole. From  (\ref{optim}) and (\ref{bestfit}) we obtain
\be\label{residuenum}
r_+=0.676\,\pm 0.608\, (stat) \pm 0.212 \,(syst).
\ee
Alternatively, one can  use  a different conformal mapping, {\em i.e.} a different value of the parameter $t_0$ in (\ref{z}), which allows a better accuracy in the high energy range. A  reasonable choice is $t_0=t_-$, when the physical region $(0,\, t_-)$  is mapped onto the interval (-0.518, 0) of the $z$ plane, and the pole position becomes $z(m_{B^*}^2, t_-)=-0.262$. Since the pole is closer to the origin, the systematic error at this point is now negligible already for the best fit with $N=4$ parameters, when we obtain
\be\label{residuenum1}
r_+= 0.544\,\pm 0.165\, (stat) \pm 0.034 \,(syst).
\ee

The value (\ref{residuenum})  can be converted to a prediction for $g_{B^*B\pi}= 2 r_+  m_{B^*} /f_{B^*}$. Using, for instance,  $f_{B^*}= 0.196 \pm 0.031\, {\rm GeV}$ \cite{Becirevic:1999}, we obtain 
\be\label{gBBpi}g_{B^*B\pi}=   37. \pm 33.\, (stat) \pm 12.\, (syst)\pm 6. (\delta f_{B^*}),
\ee
to be compared with the lattice result  $g_{B^*B\pi}=47 \pm  3\,{\rm (stat)} \pm 9\,{\rm (syst)}$  from \cite{Abada:2003}.
The large statistical and systematic errors   show that a reliable extraction of the residue from the extrapolation of our best fit is not possible. Additional information on the behavior of $f_+(q^2)$ outside the physical region, like the absence of zeros, expected on general grounds for form factors \cite{Leutwyler:2002}, or the monotony, valid in some models \cite{Isgur:1990}, might improve the prediction.  A more detailed study of this problem will be presented in a future work.

Before ending this section, let us make a few more comments on the
standard analytic parametrization (\ref{AGRS}). We presented the
results of fits to this parametrization in (\ref{N3t})-(\ref{N5t}) and
in Fig.~\ref{fig:newvstrad}. As discussed in Sec. \ref{sec:other},
the parametrization (\ref{AGRS}) has a fake singularity at the
unitarity threshold, $q^2=t_+$, which is expected to produce
distortions in the behavior of the form factor at large values of
$q^2$. For instance, from the fit with $N=5$ parameters, {\em i.e.}
using four terms in the expansion (\ref{AGRS}) and the best values
from (\ref{N5t}), we obtain 
\be f_+(t_-)=8.59\,^{+ 3.67}_{-3.55}
\,(stat) \pm 1.89 \, (syst),\quad[{\rm Eqs.  (\ref{AGRS}),
  (\ref{N5t}) }], \ee and the residue \be\label{residuenumt}
r_+= 0.956\,\pm 1.855 \, (stat) \pm 2.140 \,(syst), \quad[{\rm
  Eqs.(\ref{AGRS}), (\ref{N5t})}].  \ee 
The larger systematic errors
are explained in part by the fact that now the expansion has only four
terms (unlike in (\ref{optim}), where an additional term was
introduced using the threshold condition). The statistical errors are
also larger than for the $N=5$ fit which uses our new parametrization,
Eqs. (\ref{statsyst}) and (\ref{residuenum}), showing that the singularity at threshold  affects the behavior of the form factor near this point.

\section{Determination of $|V_{ub}|$}\label{sec:Vub}  

As shown in Sec. \ref{sec:fits}, $|V_{ub}|$ is one of the
parameters determined by our fit: the optimal value and the
statistical error are given in \eq{N5}. We chose the parametrization
such that the systematic error can be neglected along the whole physical
region. Therefore, the determination of $|V_{ub}|$ will be free of
systematic uncertainties.  Adding an experimental error of $0.01
\times 10^{-3}$ associated with the uncertainty in the $B^0$ lifetime
\cite{Barberio:2007cr}, our final prediction is
\be\label{Vub}   |V_{ub}|=(3.54 \pm 0.30) \times 10^{-3}. \ee
 This result depends of course on the theoretical and experimental input used, and will become more and more accurate as this input will improve. Our
purpose in this work was mainly to prove the advantages of the simple 
 parametrization \eq{optim} of the form factor, which we recommend as a
useful tool in future data analyses.

The result (\ref{Vub}) is consistent with the most recent prediction
from exclusive $B\to\pi$ decays \cite{Flynn:2007ii}. However, as
discussed in Sec. \ref{sec:other}, the analysis in
\cite{Flynn:2007ii} is based on a parametrization that does not fully
satisfy the constraints of analyticity and unitarity. Moreover, the
statistical correlations in the lattice results are neglected
there. Thus, our analysis puts the extraction of $|V_{ub}|$ from
exclusive $B\to\pi$ decays on a more rigorous basis.
 
From the fit given in (\ref{N5}) we  obtain also
\be\label{Vubf0}
|V_{ub}| f_+(0)=    (8.99 \pm 0.72 ) \times 10^{-4}\,,
\ee
to be compared with the result $|V_{ub}| f_+(0)= (7.6 \pm 1.9) \times
10^{-4} $ obtained with SCET and factorization \cite{Bauer:2004tj}.

\section{Conclusions}

We proposed a simple analytic parametrization for the semileptonic
$B\to\pi$ vector form factor $f_+(q^2)$, by multiplying the factor
accounting for the $B^*$ pole with a convergent expansion in powers of
a conformal mapping variable. The parametrization has the correct
behavior at the unitarity threshold and satisfies perturbative
scaling and the constraint derived from the positivity of the
correlation function of the $\bar u\gamma_\mu b$ current and its
Hermitian conjugate.  The latter was used also to define the
systematic error due to the truncation of the expansion. By increasing
up to $K=4$ the number of terms in the expansion, we obtained the
representation given in Eqs. (\ref{optim})-(\ref{bestfit}), where the
systematic error can be neglected along the whole physical
region. From the combined fit of our parametrization to experimental
results for the differential decay rate and to theoretical results for
the form factor, we obtained a prediction for $|V_{ub}|$ given in
(\ref{Vub}).  Our result confirms that $ |V_{ub}|$ extracted from the
exclusive $B\to \pi$ decays is consistent with the global fits of the
CKM matrix \cite{Charles:2004jd}.

\begin{acknowledgments}
The authors thank J. Charles for useful discussions. This work
was conducted within the framework of the Cooperation Agreement
between the CNRS and the Romanian Academy (Project CPT-Marseille -
NIPNE Bucharest), with support from the EU RTN Contract No.
MRTN-CT-2006-035482 (FLAVIAnet), from the CNRS's GDR Grant No. 2921
(``Physique subatomique et calculs sur r\'eseau'') and from the
Program Corint/ATLAS, of Romanian ANCS.
\end{acknowledgments}


\appendix*

\section{}

In this Appendix we show  that unitarity allows one to derive a bound on
the remainder of the expansion (\ref{new}), defined as
\be\label{remainder}
\delta f_+(q^2)= \frac{1}{1-q^2/m_{B^*}^2}\,\sum\limits_{K+1}^{\infty}
b_k z^k\,.
\ee 
For simplicity we omit the dependence of the coefficients and the
variable $z$ upon $t_0$, which is kept fixed in the expressions given
below.

Using (\ref{AGRS}) and (\ref{new}) we express each coefficient $b_k$
as
\be\label{btoa}
b_k=\sum\limits_{j=0}^k \tilde{\eta}_{k-j} a_j, \quad\quad k\ge 0 \,,
\ee
where $ \tilde{\eta}_j  $ appear in the expansion
\be\label{phiinvexpan}
1/\Psi(z) =\sum\limits_{k=0}^\infty \tilde{\eta}_j z^j \,,
\ee
with $\Psi(z)$ given in \eq{Psi}.
By the Cauchy inequality we obtain from (\ref{btoa})
\be\label{Cauchy}
|b_k|\le \{\sum\limits_{j=0}^k \tilde{\eta}_j^2 \, \sum\limits_{j=0}^k a_j^2 
\}^{1/2}, \quad\quad k\ge 0 \,,
\ee
and, using (\ref{unitak}),
\be\label{boundbk}
|b_k|\le \{\sum\limits_{j=0}^k \tilde{\eta}_j^2 \}^{1/2}, \quad\quad k\ge 0  \,.
\ee
Therefore, the remainder (\ref{remainder}) is bounded in terms of
calculable quantities
\be\label{tr2}\hspace{-0.05cm} |\delta f_+(q^2)| \le\, \frac{1}{|1-q^2/m_{B^*}^2|}
\sum\limits_{k=K+1}^{\infty} \{\sum\limits_{j=0}^k
\tilde{\eta}_j^2\}^{1/2}\,|z|^k\,.  \ee
The upper bound (\ref{tr2}) can be made sufficiently small for a
certain $K$ and $|z|<1$. This follows from the properties of the
coefficients $\tilde\eta_j$ defined in (\ref{phiinvexpan}): indeed,
the function $1/\Psi(z)$ has singularities on the boundary $|z|=1$,
but it is analytic inside the disk $|z|<1$.  Therefore, although the
Taylor coefficients $\tilde\eta_j$ increase with $j$, the increase is
such that sum $\sum_{k> K} \tilde\eta_k |z|^k$ can be made
arbitrarily small for a certain $K$ and $|z|<1$. The same is
true for the coefficients appearing in (\ref{tr2}), this fact being
obvious, in particular, if we use the upper bound
$\{\sum\limits_{j=0}^k \tilde{\eta}_j^2\}^{1/2} < k \tilde{\eta}_k$,
valid for sufficiently large $k$. Using this estimate, the sum in
(\ref{tr2}) is related to the remainder of the Taylor expansion of the
derivative of the function $1/\Psi(z)$, which can be made arbitrarily
small since the series is convergent for $|z|<1$.


\end{document}